\documentclass[reprint,prl,aps]{revtex4-1}
\usepackage{graphicx}
\usepackage{color}
\usepackage{dcolumn}
\usepackage{bm}
\usepackage{amssymb}
\usepackage{color,amsmath}
\usepackage{soul,xcolor} 
\setstcolor{red} 
\usepackage{epstopdf}
\usepackage[colorlinks=true, allcolors=blue]{hyperref}
\usepackage{cleveref}
\usepackage{lineno}

\newcolumntype{C}[1]{>{\centering\arraybackslash}m{#1}}
\newcolumntype{N}{@{}m{0pt}@{}}

\newcommand{\CVS}{CsV$_3$Sb$_5$}
\newcommand{\Tcdw}{$T_{\rm CDW}$}

\begin{document}

\title{Absence of $E_{2g}$ nematic instability and dominant $A_{1g}$ response in the kagome metal CsV$_3$Sb$_5$}

\author{Zhaoyu Liu$^{1}$}
\author{Yue Shi$^{2}$}
\author{Qianni Jiang$^{1}$}
\author{Elliott W. Rosenberg$^{1}$}
\author{Jonathan M. DeStefano$^{1}$}
\author{Jinjin Liu$^{3,4}$}
\author{Chaowei Hu$^{1}$}
\author{Yuzhou Zhao$^{1,2}$}
\author{Zhiwei Wang$^{3,4}$}
\author{Yugui Yao$^{3,4}$}
\author{David Graf$^{5}$}
\author{Pengcheng Dai$^{6}$}
\author{Jihui Yang$^{2}$}
\author{Xiaodong Xu$^{1,2}$}
\author{Jiun-Haw Chu$^{1}$}

\affiliation{$^{1}$Department of Physics, University of Washington, Seattle, Washington, 98195, USA}
\affiliation{$^{2}$Department of Materials Science and Engineering, University of Washington, Seattle, Washington, 98195, USA}
\affiliation{$^{3}$Centre for Quantum Physics, Key Laboratory of Advanced Optoelectronic Quantum Architecture and Measurement (MOE), School of Physics, Beijing Institute of Technology, Beijing 100081, China}
\affiliation{$^{4}$Beijing Key Lab of Nanophotonics and Ultrafine Optoelectronic Systems, Beijing Institute of Technology, Beijing 100081, China}
\affiliation{$^{5}$National High Magnetic Field Laboratory, Florida State University, Tallahassee, FL, 32306, USA}
\affiliation{$^{6}$Department of Physics and Astronomy, Rice University, Houston, Texas 77005, USA}


\maketitle

\textbf{Ever since the discovery of the charge density wave (CDW) transition in the kagome metal CsV$_3$Sb$_5$, the nature of its symmetry breaking is under intense debate. While evidence suggests that the rotational symmetry is already broken at the CDW transition temperature ($T_{\rm CDW}$), an additional electronic nematic instability well below $T_{\rm CDW}$ has been reported based on the diverging elastoresistivity coefficient in the anisotropic channel ($m_{E_{2g}}$). Verifying the existence of a nematic transition below $T_{\rm CDW}$ is not only critical for establishing the correct description of the CDW order parameter, but also important for understanding low-temperature superconductivity. Here, we report elastoresistivity measurements of CsV$_3$Sb$_5$ using three different techniques probing both isotropic and anisotropic symmetry channels. Contrary to previous reports, we find the anisotropic elastoresistivity coefficient $m_{E_{2g}}$ is temperature-independent, except for a step jump at $T_{\rm CDW}$. The absence of nematic fluctuations is further substantiated by measurements of the elastocaloric effect, which show no enhancement associated with nematic susceptibility. On the other hand, the symmetric elastoresistivity coefficient $m_{A_{1g}}$ increases below $T_{\rm CDW}$, reaching a peak value of 90 at $T^* = 20$ K. Our results strongly indicate that the phase transition at $T^*$ is not nematic in nature and the previously reported diverging elastoresistivity is due to the contamination from the $A_{1g}$ channel.}

\section{I. INTRODUCTION}
\label{intro}
Kagome metals have emerged as a new platform to investigate the interplay between topology and correlation owing to their unique lattice structures~\cite{neupert_charge_2022}. The frustrated corner-sharing triangular lattice naturally gives rise to electronic structures with flat bands, van Hove singularities and Dirac crossings~\cite{guo2009topological}. Several interesting phenomena have been discovered in kagome metals, including the giant anomalous Hall effect (AHE) in Weyl semimetal Co$_3$Sn$_2$S$_2$ \cite{liu_giant_2018}, massive Dirac fermions in Fe$_3$Sn$_2$ \cite{ye2018massive}, and charge density wave (CDW) order in FeGe, ScV$_6$Sn$_6$ and the \textit{A}V$_3$Sb$_5$ (\textit{A} = K, Rb, Cs) family~\cite{arachchige2022charge, teng2022discovery, ortiz_cs_2020}. Among the kagome metals, the \textit{A}V$_3$Sb$_5$ (\textit{A} = K, Rb, Cs) family has attracted significant attention due to the exotic behavior of its CDW phase ($T_{\rm CDW}$ = 78$\sim$104 K) and superconducting phase ($T_{\text{c}}$ = 1$\sim$3 K)~\cite{ortiz_new_2019,ortiz_cs_2020, liang_three-dimensional_2021,zhao2021cascade,jiang2021unconventional,  li2022discovery, xiao_coexistence_2023, jiang_kagome_2023, li_observation_2021, xie_electron-phonon_2022, christensen_theory_2021, wilson2024av}. In particular, the nature of the symmetry breaking in the CDW phase is not yet settled despite extensive investigations. Early studies including the measurements of AHE, optical Kerr effect, and change of muon relaxation rate all suggested that time-reversal symmetry is broken in the CDW phase \cite{yang_giant_2020, yu_concurrence_2021, yu_evidence_2021, xu_three-state_2022, mielke_time-reversal_2022, chen2022anomalous, guo2022switchable,xing2024optical}, raising the intriguing possibility that the CDW is a form of loop current order \cite{feng_low-energy_2021, christensen_loop_2022-1}. 
Nevertheless, these observations are challenged by the most recent measurement of high-resolution polar Kerr effect, which found no observable Kerr response in zero field~\cite{saykin2022high, wang2024resolving}. Therefore, whether time-reversal symmetry is truly broken remains an open question.

\begin{figure*}[t]
\includegraphics[width=7 in]{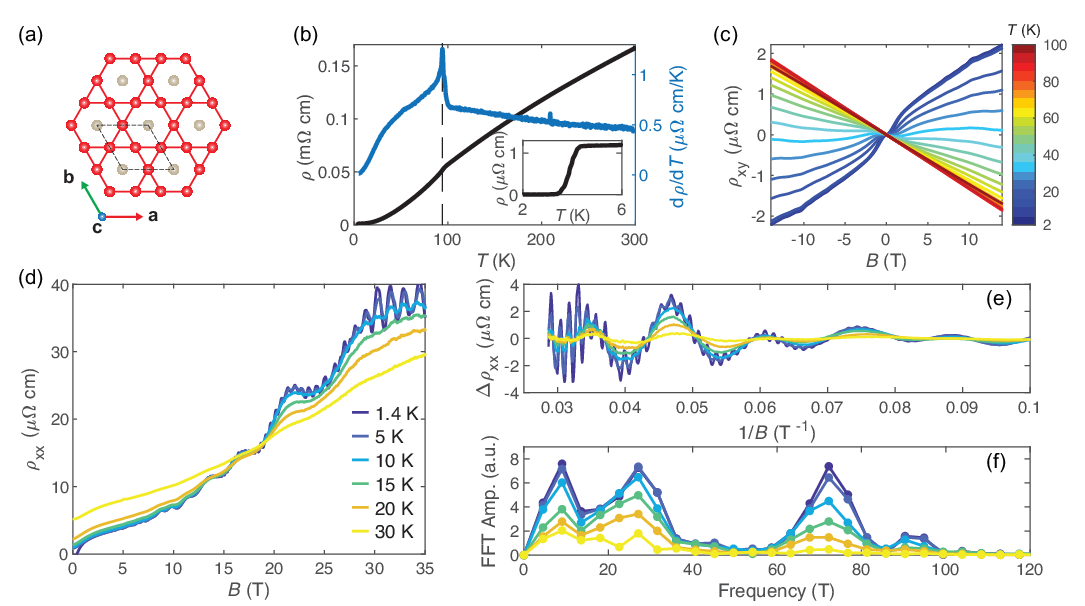} 
\caption{Electric transport characterization of CsV$_3$Sb$_5$.
(a) In-plane kagome net composed of vanadium atoms with antimony atoms filled in the center of the hexagonal lattice. 
(b) Temperature dependence of in-plane resistivity and its first derivative. The CDW transition temperature $T_{\rm CDW} = 94$ K is shown as a vertical dashed line. The inset shows the superconducting transition at 4 K. The residual resistivity ratio is 143.5. 
(c) Magnetic field dependence of Hall resistivity at different temperatures with the field direction along the $c$ axis.
(d) Magnetoresistance below 30 K. 
(e) The polynomial background-subtracted magnetoresistance in (d) plotted as a function of the inverse magnetic field. 
(f) The fast Fourier transform of the quantum oscillations in (e). 
}
\label{fig:1}
\end{figure*}

In addition to time-reversal symmetry, the rotational symmetry in the CDW phase has also been intensely studied. A two-fold anisotropy in the CDW phase that breaks the $C_6$ rotational symmetry was first reported by angle-dependent magnetoresistance measurements~\cite{xiang_twofold_2021}. Further studies including scanning tunneling spectroscopy (STM) \cite{li_rotation_2022, li_unidirectional_2023}, Raman spectroscopy \cite{wulferding_emergent_2022,wu_charge_2022}, optical birefringence measurements \cite{wu_simultaneous_2022, xu_three-state_2022}, and angle-resolved photoemission spectroscopy (ARPES) \cite{jiang_observation_2023, luo2022electronic, kang2023charge}, all confirmed the observed two-fold anisotropy. However, the temperature at which rotational symmetry is broken remains a key question. While most studies agree that the breaking of rotational symmetry happens at \Tcdw, a diverging elastoresistivity response was reported within the CDW phase \cite{nie_charge-density-wave-driven_2022,sur2023optimized}, which was argued as evidence for an electronic nematic instability well below $T_{\rm CDW}$.

The electronic nematic phase refers to a spontaneous rotational-symmetry-breaking phase while preserving translational symmetry, which has been extensively studied in strongly correlated electronic systems \cite{fradkin2010nematic, fernandes2014drives}. In the context of Fe-based superconductor \cite{chu2012divergent, kuo2016ubiquitous, liu2016nematic, gu2017unified}, nematicity is understood as a vestigial order of the underlying spin density wave or CDW phases \cite{nie2014quenched,fernandes2014drives,fernandes2019intertwined, bohmer2022nematicity}. The partial melting of the density waves destroys long-range periodicity while continuing to break rotational symmetry, hence the nematic transition is always above the density wave transition in these materials. A salient feature of the nematicity is a diverging nematic susceptibility at temperatures above the phase transition , which can be probed by elastoresistivity measurements, where resistivity anisotropy serves as a proxy for the nematic order parameter and the anisotropic strain serves as its conjugate field. For instance, a diverging elastoresistivity coefficient in the anisotropic strain channel ($m_{B_{2g}}$) with a Curie-Weiss temperature dependence was observed above the nematic transition in Ba(Fe$_{1-x}$Co$_x$)$_2$As$_2$~\cite{chu2012divergent}.

\begin{figure*}[t]
\includegraphics[width=7 in]{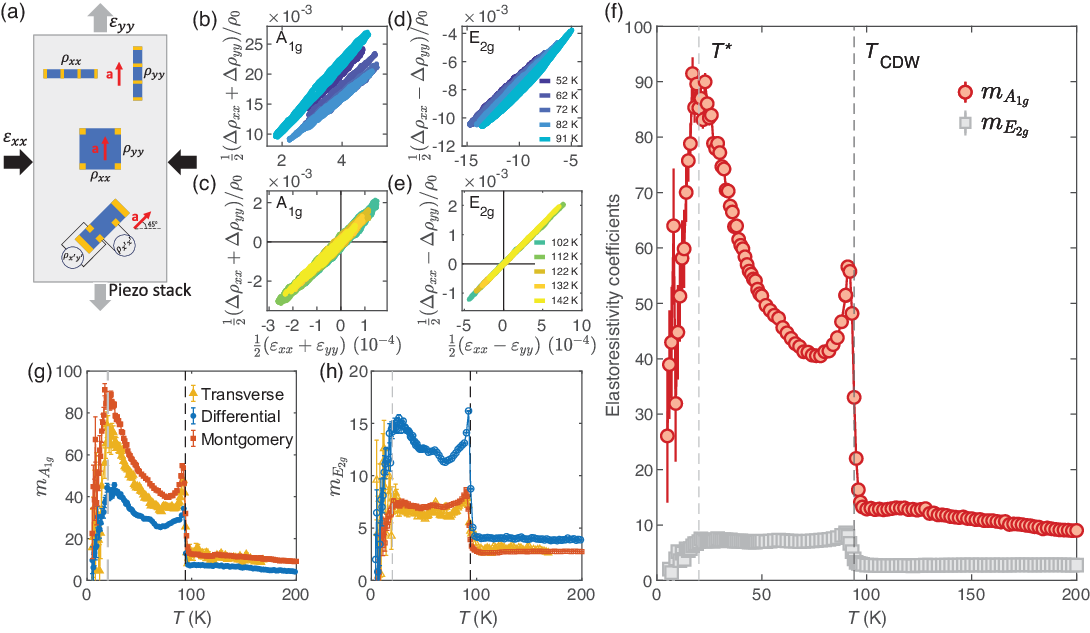} 
\caption{Measurements of elastoresistivity coefficients in \CVS. (a) Schematic of three elastoresistivity measurement techniques. Samples are glued on the side wall of a piezostack (gray rectangular). The differential technique utilizes two bar-shaped samples, glued orthogonal to each other (top), the modified Montgomery technique uses a single square sample, with four contacts on the corner (middle), and the transverse technique uses on a five-contact bar-shaped sample rotated 45 degrees with respect to the poling direction of the piezostack (bottom). The red arrow shows the crystal axis for each configuration. 
(b-e) Representative data of the resistivity as a function of strain in the isotropic $A_{1g}$ and anisotropic $E_{2g}$ channels at several temperatures measured by the modified Montgomery technique. 
(f) The temperature dependence of elastoresistivity coefficients measured by the modified Montgomery technique. Black dashed line represents \Tcdw~and the gray dashed line indicates $T^*$.
(g) The temperature dependence of  $m_{A_{1g}}$ measured by three techniques. 
(h) Same plot as (g) for $m_{E_{2g}}$. 
}
\label{fig:2}
\end{figure*}

In this context, a diverging elastoresistivity in CsV$_3$Sb$_5$ that peaks at $T^* \approx 35$ K, a temperature well below $T_{\rm CDW}$, is rather unusual~\cite{nie_charge-density-wave-driven_2022}. It would imply the rotational symmetry is not broken at \Tcdw, which is inconsistent with the majority of previous experimental observations \cite{li_unidirectional_2023,wu_simultaneous_2022,xu_three-state_2022}. Another possible explanation is that the rotational symmetry is only weakly broken at $T_{\rm CDW}$ by the $\pi$ phase shift of the CDW order between different kagome layers \cite{park2021electronic}, while the CDW within the two-dimensional kagome plane remains isotropic. At a lower temperature $T^*$, the rotational symmetry within the kagome plane is broken, leading to diverging elastoresistivity~\cite{grandi2023theory}. In either scenario, the existence of a nematic instability below \Tcdw\ provides strong constraints to distinguish between various proposals for the CDW order parameter~\cite{park2021electronic,christensen_theory_2021,denner2021analysis, grandi2023theory,tazai2022mechanism,christensen_loop_2022-1,zhou2022chern}. In addition, a very recent elastoresistivity study proposed a nematic quantum critical point residing within the first superconducting dome in the phase diagram of Ti doped CsV$_3$Sb$_5$~\cite{sur2023optimized}, which was argued as evidence for nematic fluctuation enhanced superconducting pairing. On the other hand, it has been shown that non-ideal experiment configurations, such as what is incorporated in the differential elastoresistivity technique used in reference \cite{nie_charge-density-wave-driven_2022} and \cite{sur2023optimized}, may lead to the mixing of elastoresistivity signals in different symmetry channels, resulting in a false diverging nematic response even though the leading instability is in other symmetry channels \cite{wiecki2020dominant, wiecki2021emerging, rosenberg2019divergence, ye2023elastocaloric}. The purpose of this work is to perform a comprehensive study of elastoresistivity and the elastocaloric effect in CsV$_3$Sb$_5$ to elucidate the symmetry associated with the proposed transition at $T^*$.

\section{II. RESULTS}
\subsection{A. Electrical transport characterization}

\CVS~single crystals have a hexagonal shape with the $a$ axis along the natural growth edges, consistent with the six-fold rotational symmetry of the kagome lattice [Fig. \ref{fig:1}(a)]. The CDW transition occurs at $T_{\rm CDW} = 94$ K, which can be seen from a resistivity anomaly and a sharp peak in $\mathrm{d}\rho/\mathrm{d}T$ as shown in Fig. \ref{fig:1}(b).  The superconducting transition is at $T_{\text{c}} = 3$ K. The residual resistivity ratio of $\rho(300~\rm{K})/\rho (5~\text{K}) = 143.5$ indicates the high quality of the single crystals. The AHE is observed in the in-plane Hall resistivity below $T_{\rm CDW}$ [Fig. \ref{fig:1}(c)], consistent with previous experiments \cite{yu_concurrence_2021}. The Shubnikov–de Haas quantum oscillations (QOs) can be observed in longitudinal resistivity below 30 K [Fig. \ref{fig:1}(d)]. The oscillatory component $\Delta\rho_{xx}$ as a function of inverse field and its fast Fourier transform (FFT) are shown in Fig. \ref{fig:1}(e, f), respectively. Four principal frequencies at 11, 28, 73, and 90 T were observed in the low-frequency regime (with some high-frequency peaks not shown here) in high field measurements up to 35 T. These observations are all in agreement with previous reports, providing a solid foundation to further study the elasto-response in these single crystals \cite{yu_concurrence_2021, ortiz_fermi_2021, fu2021quantum}.

\subsection{B. Elastoresistivity}

Elastoresistivity is a fourth-rank symmetric tensor that relates the change of resistivity of a system to the externally induced strains. When using Voigt notation, the elastoresistivity tensor can be expressed as a second-rank $6\times6$ tensor:
\begin{equation} \label{eq1}
    m_{ij}=\frac{\partial(\Delta\rho/\rho)_i}{\partial\varepsilon_j}
\end{equation}
where the indices {$i,j=1-6$} represent $1 = xx$, $2 = yy$, etc. \cite{kuo2016ubiquitous}. These elastoresistivity coefficients can be further grouped into different symmetry channels based on the irreducible representations of the point group of the crystal lattice. For example, in $D_{6h}$, the elastoresistivity coefficients associated with the isotropic $A_{1g}$ irrep and the anisotropic $E_{2g}$ irrep are $m_{A_{1g}}=m_{11}+m_{12}-m_{13}\frac{2\nu_{ac}}{1-\nu_{ab}}$ and $m_{E_{2g}} = m_{11}-m_{12}$, respectively. Here, the $\nu_{ab, ac}$ are the in-plane and out-of-plane Poisson's ratios (see Methods\ref{methods}) \cite{fernandes2020nematicity, nie_charge-density-wave-driven_2022}. In a material with an electronic nematic instability, the diverging susceptibility associated with the nematic transition will manifest in a diverging temperature dependence of the elastoresistivity coefficient in the anisotropic symmetry channel.

When measuring elastoresistivity, single crystals are glued on the sidewall of a piezostack, which induces a combination of purely anisotropic strain $\frac{1}{2}(\varepsilon_{xx}-\varepsilon_{yy})$ and isotropic strain $\frac{1}{2}(\varepsilon_{xx}+\varepsilon_{yy})$ when an external voltage is applied to the stack. In order to separate the isotropic and anisotropic symmetry channels, both $\rho_{xx}$ and $\rho_{yy}$ need to be measured. Three experimental techniques have been developed to measure elastoresistivity coefficients. The first technique that was developed is the differential elastoresistivity technique \cite{chu2012divergent}, which measures $\rho_{xx}$ and $\rho_{yy}$ using two separate bar-shaped samples cut along the same crystal axes but oriented perpendicularly on the stack, as shown in the top configuration in Fig. \ref{fig:2}(a). However, it was soon realized that this technique inevitably introduces cross-contamination between different symmetry channels \cite{wiecki2020dominant, wiecki2021emerging}. Exact symmetry decomposition requires identical strain transmission in both samples, which is never the case in any practical experiment. In fact, the bar-shaped sample results in  more effective strain transmission for the uniaxial strain along the bar direction. Therefore, a bar-shaped sample glued along the $x (y)$ direction experiences dominantly uniaxial strain $\varepsilon_{xx} (\varepsilon_{yy})$ even though nominally the same anisotropic strain $\varepsilon_{xx} - \varepsilon_{yy}$ was applied. To address this issue, the modified Montgomery technique \cite{kuo2016ubiquitous} and the transverse method \cite{shapiro2016measurement} were subsequently developed. The modified Montgomery technique allows for obtaining $\rho_{xx}$ and $\rho_{yy}$ using a single square sample [middle configuration in Fig. \ref{fig:2}(a)] and the transverse technique enables the direct measurement of the resistivity anisotropy $\rho_{x'y'}$ in a five-contact bar-shaped sample [bottom configuration, see details in Methods\ref{methods}]. Both methods measure the full resistivity tensors from the same single crystalline samples, hence they do not suffer cross-contamination issues, and the symmetry decomposition is exact.

To thoroughly examine the elastoresistivity coefficients of \CVS, all three techniques mentioned above were employed. We found that resistivity as a function of strain is linear in both $E_{2g}$ and $A_{1g}$ symmetry channels at all temperatures for all measurements [Fig. \ref{fig:2}(b-e)], with only weak hysteresis near $T_{\rm CDW}$ likely due to structural domains. This suggests that all the measured elastoresistivity coefficients are in the near-zero strain linear response regime. Figure \ref{fig:2}(f) shows the temperature dependence of $E_{2g}$ and $A_{1g}$ elastoresistivity coefficients measured by the modified Montgomery technique. The $m_{E_{2g}}$ value jumps from 3 to 8 at $T_{\rm CDW}$, but it is essentially temperature independent both below and above $T_{\rm CDW}$. The $m_{A_{1g}}$ is also temperature independent above $T_{\rm CDW}$, albeit with a larger value ($\approx$ 10). At $T_{\rm CDW}$ it exhibits a sharp peak and grows continuously as temperature decreases until reaching a maximum value of 90 at $T^* = 20$ K.

The $m_{E_{2g}}$ measured by the modified Montgomery technique is very different from those reported in Refs. \cite{nie_charge-density-wave-driven_2022} and \cite{sur2023optimized}, which were measured by the differential elastoresistivity technique. The $m_{E_{2g}}$ reported in Refs. \cite{nie_charge-density-wave-driven_2022} and \cite{sur2023optimized} is considerably larger, and shows a temperature dependence that resembles $m_{A_{1g}}$ measured by the modified Montgomery technique. To gain more insight, we present the elastoresistivity coefficients measured by all three techniques in Fig. \ref{fig:2}(g, h). It can be seen that $m_{E_{2g}}$ and $m_{A_{1g}}$ measured by the modified Montgomery technique and the transverse technique are consistent with each other, whereas the differential elastoresistivity technique yields a larger $m_{E_{2g}}$ and smaller $m_{A_{1g}}$ in comparison to the other two techniques. The temperature dependence of $m_{E_{2g}}$ measured by the differential elastoresistivity technique is also more similar to that of $m_{A_{1g}}$. All of these are consistent with the admixture of $m_{A_{1g}}$ into $m_{E_{2g}}$ in the differential elastoresistivity measurement due to unequal strain transmission in the two samples. Hence, we conclude that the divergent $m_{E_{2g}}$ in previous reports is not intrinsic.
We note that two groups have also reported elastoresistivity measurements during the preparation of this manuscript, which are in broad agreement with our observations \cite{frachet2024colossal, asaba2024evidence}. A detailed comparison of the results among different groups is discussed in the Supplementary Materials.

\subsection{C. Elastocaloric effect}

\begin{figure}[t]
\includegraphics[width=3.4in]{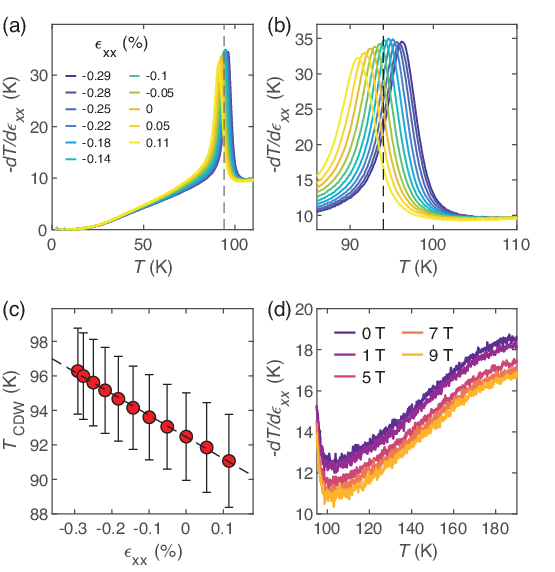} 
\caption{AC Elastocaloric effect.
(a) Temperature dependence of the ECE measured at different DC bias uniaxial strains. The uniaxial stress was applied along the longitudinal direction of a bar-shaped sample cut along the $a$ axis. The data was taken at a 1 K/min cooling rate. 
(b) A zoom-in view of the peaks in ECE near $T_{\rm CDW}$. 
(c) The CDW transition temperature as a function of strain determined by the peaks in ECE at $T_{\rm CDW}$. The error bar is determined by the full width at half maximum of the peaks. The dashed line represents a linear fit to $T_{\rm CDW}(\varepsilon)$ with a slope of $dT_{\rm CDW}/d\varepsilon_{xx}=-12.6 ~$K$/\%$. 
(d) ECE data above the CDW transition under various magnetic fields measured in  a different sample. The magnetic field is along the $c$ axis, and the DC-biased strain is near zero. There is no change of ECE signal as a function of magnetic field except for a slight shift of the background due to the weak field dependence of the voltage response of a type-E thermocouple above 100 K \cite{inyushkin1998magnetic}.
}
\label{fig:3}
\end{figure}

In addition to the elastoresistivity measurements, the elastocaloric effect (ECE) is another sensitive probe to measure the diverging susceptibility associated with a nematic phase transition \cite{ikeda2019ac,ikeda2021elastocaloric}. The elastocaloric measurement detects the temperature change of a system due to an adiabatically induced strain. Experimentally, a small AC strain is induced in the sample simultaneously with a tunable DC bias strain, and the AC temperature variation is measured at the frequency of the induced strain by a thermocouple (see details in Methods). 
The elastocaloric coefficient ($\frac{\partial T}{\partial\varepsilon}$) can be related to the isothermal entropy change caused by strain via the following equation:  
\begin{equation}
    \left( \frac{\partial T}{\partial\varepsilon} \right)_S = -\frac{T}{C_{\varepsilon}} \left( \frac{\partial S}{\partial\varepsilon} \right)_{T}
\end{equation}
where $C_{\varepsilon}$ is the heat capacity at a given strain and $S$ is entropy. The elastocaloric effect is sensitive to both the susceptibility of the nematic order parameter as well as its spontaneous onset. Depending on the symmetry of the induced strain, two ECE phenomena are expected near a phase transition~\cite{hristov2019elastoresistive,ikeda2021elastocaloric}. 
The first of which is associated with inducing strain that breaks the same symmetry as the order parameter \cite{ikeda2021elastocaloric}. When using such a strain to perturb the system, an enhancement of ECE above the transition temperature is expected. This enhanced ECE is described by the following equation: 
\begin{equation}
    \left( \frac{\partial T}{\partial\varepsilon_i} \right)_S = -\frac{T\lambda^2\varepsilon_i}{C_{\varepsilon_i}} \left( \frac{d \chi_N}{dT} \right)
\end{equation}
where $i$ labels the anisotropic symmetry channels. Thus, this enhancement of the ECE is proportional to both the temperature derivative of the susceptibility ($\frac{d \chi_N}{dT}$) and the DC bias anisotropic strain $\varepsilon_i$, hence it switches sign from tensile to compressive DC strain. This effect was observed in the iron-based superconductors where $\varepsilon_{B_{2g}}$ couples linearly to electronic nematicity and showed excellent agreement with the nematic susceptibility obtained from previous elastoresistivity measurements \cite{ikeda2021elastocaloric, straquadine2020frequency}.

The second effect is associated with the temperature shift of the phase transition induced by strains with symmetry that do NOT couple linearly to the order parameter. For example, $\varepsilon_{A_{1g}}$ is expected to linearly tune $T_{\rm CDW}$. In this case we expect the ECE near the phase transition to be proportional to the critical contribution of heat capacity ($C^{(c)}_{\varepsilon_{E_{2g}}}$) times the strain derivative of transition temperature:

\begin{equation}
    \left( \frac{\partial T}{\partial\varepsilon_{A_{1g}}} \right)_S = ~\frac{C^{(c)}_{\varepsilon_{E_{2g}}}}{C_{\varepsilon_{E_{2g}}}} \frac{d T_{\rm CDW}}{d \varepsilon_{A_{1g}}}
\end{equation}

We measured the ECE of a \CVS\ sample by applying a uniaxial stress, which induces both $\varepsilon_{A_{1g}}$ and $\varepsilon_{E_{2g}}$. Figure \ref{fig:3}(a) summarizes the ECE measured under different DC bias strains. A pronounced peak that mimics the heat capacity anomaly is observed near $T_{\rm CDW}$, and the peak is systematically shifted as a function of DC bias strain. This peak is consistent with the second effect mentioned above, where $\varepsilon_{A_{1g}}$ linearly shifts the transition temperature. The linear dependence of $T_{\rm CDW}$ as a function of $A_{1g}$ strain is also consistent with previous studies of \CVS ~\cite{qian2021revealing}.

\begin{figure}[b]
\includegraphics[width=3 in]{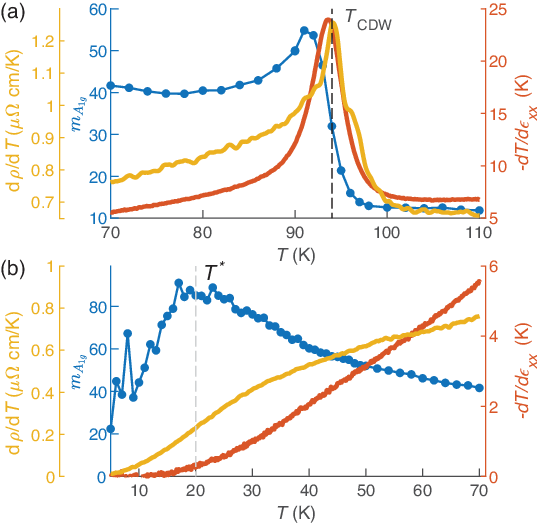}
\caption{Comparison of $m_{A_{1g}}$, ECE and $d\rho/dT$ in \CVS. Temperature dependence of $m_{A_{1g}}$ (blue points), $-dT/d\varepsilon_{xx}$ at zero strain (red line), and $d\rho/dT$ (yellow line) are plotted near $T_{\rm CDW}$ (vertical dashed line), (a), and in the low-temperature range (b). The vertical gray dashed line indicates $T^*$. 
}
\label{fig:4}
\end{figure}

\begin{figure*}
\includegraphics[width=6.5 in]{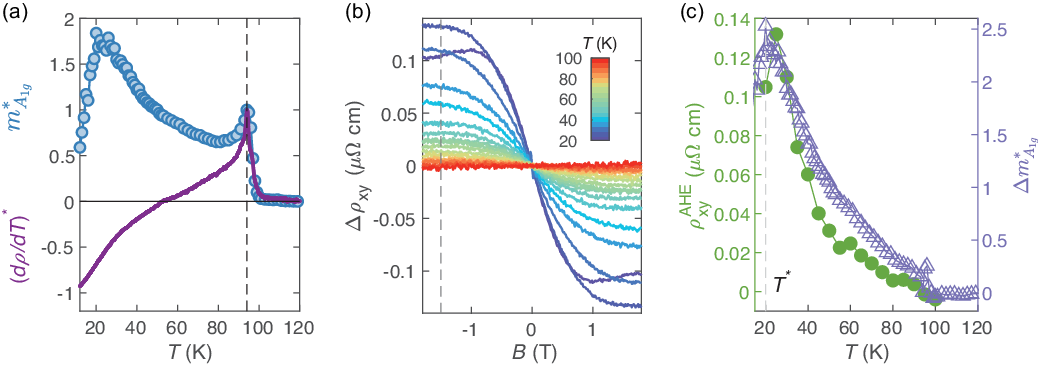} 
\caption{Similarity between the temperature dependence of AHE and $m_{A_{1g}}$.
(a) Temperature dependence of elastoresistivity $m_{A_{1g}}$ and the temperature derivative of resistivity normalized by the peak value at \Tcdw. The elastoresistivity $m_{A_{1g}}$ was subtracted by a temperature-independent background (its value at 120 K, $m_{A_{1g}}^{120 \rm K}$), and normalized by its peak value at \Tcdw, i.e., $m_{A_{1g}}^*=(m_{A_{1g}}-m_{A_{1g}}^{120 \rm K})/(m_{A_{1g}}^{T_{\rm CDW}}-m_{A_{1g}}^{120 \rm K})$. The same procedure was applied to $d\rho/dT$ to obtain $(d\rho/dT)^*$. We shifted $m_{A_{1g}}^*$ by 3 K because of the change of \Tcdw~due to thermal built-in strain.
(b) $\Delta\rho_{xy}$ extracted by subtracting the linear ordinary Hall background at various temperatures (see details in Supplementary Materials). The black dashed line is located at -1.5 T, where we extracted the $\rho_{xy}^{\rm{AHE}}$ values. 
(c) $\rho_{xy}^{\rm{AHE}}$ and $\Delta m_{A_{1g}}^*$ as a function of temperature, where $\Delta m_{A_{1g}}^* = m_{A_{1g}}^* - (d\rho/dT)^*$. This presents the temperature dependence of $m_{A_{1g}}$ without the peak feature.
}
\label{fig:5}
\end{figure*}

However, the enhancement of the ECE due to a diverging nematic susceptibility was not observed either above or below $T_{\rm CDW}$. The signal quickly converges both above and below $T_{\rm CDW}$ [Fig. \ref{fig:3}(b)], and there is no indication of concavity or slope change in the peak from tensile to compressive strains. 
This result is consistent with no diverging nematic susceptibility, further corroborating the conclusion established by the elastoresistivity measurements.
A very recent study suggested an odd-parity nematic transition above \Tcdw\, which turns into a first order transition under $c$-axis magnetic field \cite{asaba2024evidence}. We also performed ECE measurements under $c$-axis magnetic field up to 9 T and observed no anomaly above \Tcdw\, as illustrated in Fig.~\ref{fig:3}(d). This result is consistent with the absence of such a transition measured by thermal expansion experiments \cite{frachet2024colossal}.

\section{III. Discussion}

Both elastoresistivity and elastocaloric effect measurements suggest the absence of the nematic instability below \Tcdw. The diverging $m_{E_{2g}}$ reported in Refs. \cite{nie_charge-density-wave-driven_2022} and \cite{sur2023optimized} is most likely due to the mixing of the diverging $m_{A_{1g}}$, a consequence of unequal strain transmissions in the differential elastoresistivity technique. We note that in Ref. \cite{sur2023optimized}, the Montgomery elastoresistivity measurement was performed on one Ti-doped \CVS\ sample, which showed agreement with the differential elastoresistivity measurement. Given that the unequal strain transmission is a generic issue in all differential elastoresistivity measurements, a reinvestigation of the elastoresistivity of Ti doped \CVS\ is highly warranted.

We also note that our result is not inconsistent with the rotational symmetry breaking at \Tcdw. The increase of $m_{E_{2g}}$ could be a result of aligning symmetry breaking CDW domains by the anisotropic strain \cite{li_rotation_2022, xu_three-state_2022, wu2022simultaneous}, but the small value of $m_{E_{2g}}$ ($\sim$8) indicates that the electronic anisotropy is only moderately larger than the structural anisotropy \cite{guo2024correlated}, which could be consistent with the weak anisotropy introduced by the inter-layer coupling. In this scenario, given that both rotational symmetry and translational symmetry are simultaneously broken at \Tcdw, this transition cannot be characterized as a nematic transition. 
The absence of diverging {$E_{2g}$} elastoresistivity and elastocaloric effect above $T_{\rm CDW}$ also confirms that there is no {fluctuating} vestigial nematicity associated with the rotational symmetry breaking CDW. The temperature-independent $E_{2g}$ elastoresistivity also suggests that the previously reported 4$a_0$ order observed by STM is likely a surface phenomenon, which does not contribute to bulk electronic anisotropy \cite{zhao2021cascade,li2022discovery}.
Instead of diverging $m_{E_{2g}}$, the dominant response is in the isotropic $A_{1g}$ channel, manifested by the diverging $m_{A_{1g}}$. Below we discuss the possible origins of the large and strongly temperature dependent $m_{A_{1g}}$. There are two features in the temperature dependence of $m_{A_{1g}}$. The first feature is a sharp peak at \Tcdw, which can be understood by the generalized Fisher-Langer relation \cite{hristov2019elastoresistive}. The relation states that the temperature derivative of resistivity, $d\rho/dT$, the ECE, and elastoresistivity coefficient all scale like the critical component of the heat capacity near a phase transition, if the strain used in the ECE and elastoresistivity is associated with the same symmetry channel that is orthogonal to the order parameter. Indeed, as shown in Fig. \ref{fig:4}(a), all three quantities follow the same temperature dependence near \Tcdw. As a side note, this phenomenon can appear in the $m_{E_{2g}}$ channel as well, if the sample is at a nonzero $E_{2g}$ strain, since \Tcdw~is potentially a quadratic function of $\epsilon_{E_{2g}}$.

 Below \Tcdw, the $m_{A_{1g}}$ increases rapidly as temperature decreases, showing a diverging behavior that peaks at $T^*$. In contrast to the sharp peak at \Tcdw, we do not see similar temperature dependence in the ECE, and $d\rho/dT$ only shows a broad hump centered at $T^*$ that barely resembles the feature in $m_{A_{1g}}$ [Fig. \ref{fig:4}(b)]. The lack of any feature in the ECE and the strong divergence in elastoresistivity is rather striking. If we attribute the divergence of $m_{A_{1g}}$ to a phase transition at $T^*$, it implies that the order parameter has a very weak coupling to strain, such that it causes a minimum entropy change induced by strain, yet it has a very strong coupling to the conducting quasi-particles, leading to the strong divergence in elastoresistivity. This phenomenon has been observed in iron-based superconductors, where the magnitude of ECE decreases by factor of 35 as the doping approaches the nematic quantum critical point, while the elastoresistivity is enhanced by fivefold \cite{ikeda2021elastocaloric}. It was understood as a consequence of simultaneous reduction of nemato-elastic coupling and enhancement of nemato-transport coupling.

Another notable observation is that the temperature dependence of $m_{A_{1g}}$ shows a striking resemblance to that of the AHE discovered in the CDW phase of the $A$V$_3$Sb$_5$ family \cite{yang_giant_2020,liu_giant_2018,yu_concurrence_2021}. 
Since the temperature dependence of $m_{A_{1g}}$ can be decomposed into two components: a peak near \Tcdw\ that resembles $d\rho/dT$ and a diverging behavior towards $T^*$. To isolate diverging behavior towards $T^*$, we normalized both $m_{A_{1g}}$ and $d\rho/dT$ [Fig. \ref{fig:5}(a)] and subtract the latter from the former. The anomalous Hall resistivity was extracted by removing the ordinary Hall from the $\rho_{xy}$ in Fig. \ref{fig:1}(c) by linearly fitting the data for fields between 1.5 and 2 T, shown in Fig. \ref{fig:5}(b). The extracted $\rho_{xy}^{\rm AHE}$ as a function of temperature is plotted against the background-subtracted $\Delta m_{A_{1g}}^{*}$, shown in Fig. \ref{fig:5}(c). The remarkable resemblance of these transport observables highlights the impact of the $T^*$ instability on the quasi-particles at the Fermi level. We notice that the recently observed chiral transport effect in \CVS\ also shows a similar temperature dependence to AHE \cite{guo2022switchable}. Intriguingly, while the AHE and the chiral transport effect are sensitive to time-reversal symmetry and inversion symmetry breaking, respectively, the $A_{1g}$ elastoresistivity is not a direct probe of either symmetry breaking order parameters, because both resistivity and strain are even-parity operators. Future studies on the strain dependent of AHE and nonlinear transport effect may elucidate the origin of the common diverging behavior of all three transport observables. There has also been suggestion that the small Fermi pockets observed in quantum oscillations may be relevant to AHE or field-tunable chirality \cite{jiang2021unconventional, chen2022anomalous, xing2024optical, guo2022switchable}. This may also be resolved by performing strain-dependent quantum oscillation measurements.

\section{IV. Conclusion}
In summary, we investigated the isotropic and anisotropic elastoresistivity coefficients, $m_{A_{1g}}$ and $m_{E_{2g}}$, as well as the elastocaloric effect in \CVS. The lack of enhancement in elastocaloric effect and the temperature-independent $m_{E_{2g}}$ below \Tcdw\ are both consistent with the absence of nematic instability in this system. The previously reported diverging elastoresistivity $m_{E_{2g}}$ is likely due to the cross-contamination from $m_{A_{1g}}$. Both the elastoresistivity coefficient $m_{A_{1g}}$ and the ECE show a peak at \Tcdw, which is a consequence of the tuning of \Tcdw\ by $A_{1g}$ strain and it is consistent with the extended Fisher-Langer relation. In addition to the peak at \Tcdw, $m_{A_{1g}}$ also shows a diverging temperature dependence below \Tcdw\ that reaches a maximum at $T^*$, but there is no corresponding elastocaloric anomaly. Our results show that there is no nematic phase transition within the CDW phase in \CVS, and the nature of $T^*$ requires further investigation.

\section{V. Methods} \label{methods}

\textit{Transport measurements.} - Single crystals of \CVS~were synthesized using a self-flux method described elsewhere \cite{ortiz_cs_2020, wang_electronic_2021, qian2021revealing}. The electrical transport measurements were performed in DynaCool (Quantum Design, Inc.). The high magnetic field experiments were carried out in a 35 T resistive magnet at the National High Magnetic Field Laboratory in Tallahassee, FL.

\textit{Elastoresistivity tensor in $D_{6h}$ point group.} - The symmetry decomposition of elastoresistivity tensor in $D_{6h}$ point group is different from that in $D_{4h}$. There are two irreducible representations in $D_{6h}$ that are relevant to this study: isotropic $A_{1g}$ and anisotropic $E_{2g}$ symmetry. The in-plane biaxial strain employed in this study can be decomposed into these two symmetry channels: $\varepsilon_{A_{1g,1}}=\frac{\varepsilon_{xx}+\varepsilon_{yy}}{2}$, $\varepsilon_{A_{1g,2}}=\varepsilon_{zz}$ and  $\varepsilon_{E_{2g}}=(\frac{\varepsilon_{xx}-\varepsilon_{yy}}{2},~\varepsilon_{xy})$. We note that $E_{2g}$ is a two-dimensional irreducible representations and $\frac{\varepsilon_{xx}-\varepsilon_{yy}}{2}$ and $\varepsilon_{xy}$ are the two bases. The in-plane resistivity tensors can also be decomposed into these two irreducible representations: 
\begin{equation} 
\begin{split}
 &(\Delta\rho/\rho)_{A_{1g,1}} = \frac{1}{2}[(\Delta\rho/\rho)_{xx}+(\Delta\rho/\rho)_{yy}]\\
 &(\Delta\rho/\rho)_{A_{1g,2}} = (\Delta\rho/\rho)_{zz}\\
 &(\Delta\rho/\rho)_{E_{2g}}   = (\frac{1}{2}[(\Delta\rho/\rho)_{xx}-(\Delta\rho/\rho)_{yy}],~(\Delta\rho/\rho)_{xy})
\end{split}
\label{eq:ERcoeff}
\end{equation}

Therefore, if we focus on the in-plane elastoresistivity coefficients in isotropic $A_{1g}$ and anisotropic $E_{2g}$ symmetry channels are related to the elastoresistivity tensors in the Cartesian coordinate systems by the following expressions:
\begin{equation} \label{eqM1}
\begin{aligned}
    m_{A_{1g}} &= \frac{(\Delta\rho/\rho)_{A_{1g}}}{\varepsilon_{A_{1g}}}\\
    &=\frac{(m_{xx,xx}+m_{xx,yy})(\varepsilon_{xx}+\varepsilon_{yy})+2m_{xx,zz}\varepsilon_{zz}}{\varepsilon_{xx}+\varepsilon_{yy}}\\
               &= m_{xx,xx} + m_{xx,yy} - m_{xx,zz}(\frac{2\nu_{ac}}{1-\nu_{ab}}) \\
    m_{E_{2g}} &= \frac{(\Delta\rho/\rho)_{E_{2g}}}{\varepsilon_{E_{2g}}} \\
    &=\frac{(m_{xx,xx}-m_{xx,yy})(\varepsilon_{xx}-\varepsilon_{yy})}{\varepsilon_{xx}-\varepsilon_{yy}}\\
    &= m_{xx,xx} - m_{xx,yy}
\end{aligned}
\end{equation}

Since the sample is glued down to the side wall of the piezostack by the Stycast, its in-plane deformation will be constrained by the deformation of the piezostacks. Consequently, $\nu_{ab}=-\epsilon_{yy}/\epsilon_{xx}$ and $\nu_{ac}=-\epsilon_{zz}/\epsilon_{xx}$ are the Poisson's ratios of the piezostacks and the sample, respectively.
We note that because $E_{2g}$ is a two-dimensional irreducible representation, $m_{E_{2g}} = m_{xy,xy} = m_{xx,xx}-m_{xx,yy}$.  In Voigt notation, $1 = xx$, $2 = yy$, hence $m_{A_{1g}}=m_{11}+m_{12}-m_{13}\frac{2\nu_{ac}}{1-\nu_{ab}}$ and $m_{E_{2g}}=m_{11}-m_{12}$.

\textit{Elastoresistivity techniques.} -  
A tunable strain was applied to single crystal samples by gluing them on the sidewall of a piezostack (P-885.11, from PI-USA) using a thermally conductive epoxy (Stycast 2850FT with Catalyst 24LV, from Loctite). The uniaxial strain along the poling direction of the piezostack, $\varepsilon_{yy}$ [Fig. \ref{fig:2}(a)], was measured by a foil strain gauge (CEA-06-062UWA-350, from Micro-Measurements), which was glued to the other side of the piezostack. The uniaxial strain along the orthogonal direction, $\varepsilon_{xx}$, was determined by the known in-plane Poisson's ratio of the piezostack. As discussed in the main text, the elastoresistivity coefficients can be measured by three techniques: 
\begin{itemize}
    \item \textit{Differential technique}. Two bar-shaped samples are glued to the piezostack. The bar directions are aligned with the transverse and poling directions of the piezostack. Using the standard four-terminal technique, the resistivity $\rho_{xx}$ and $\rho_{yy}$ are measured separately in each sample.
    \item \textit{Modified Montgomery technique}. Four electric contacts are made at the corners of a square-shaped sample, which is glued on the piezostack with the edges aligned to the poling direction. $R_{xx}$ and $R_{yy}$ are measured by sourcing current using the contacts on one side and measuring the voltage using the contacts on the other side. The measured resistance $R_{xx}$ and $R_{yy}$ can then be converted to resistivity $\rho_{xx}$ and $\rho_{yy}$ following the procedure in Supplementary Materials.
    \item \textit{Transverse technique}. The transverse technique measures the transverse resistivity $\rho_{xy}$ induced by shear strain $\varepsilon_{xy}$, which is different from the other two techniques that measures the longitudinal resistivity $\rho_{xx}$ and $\rho_{yy}$ \cite{shapiro2016measurement}. The shear strain $\varepsilon_{xy}$ is equivalent to an anisotropic strain $\frac{\varepsilon_{xx}-\varepsilon_{yy}}{2}$ under 45 degrees rotation. In the transverse technique, a Hall bar with five electrical contacts is glued onto the piezostack and is aligned at a 45$^{\circ}$ angle with respect to the $y$ axis. Note that there is no magnetic field, hence the transverse resistivity $\rho_{xy}$ is not a result of the Hall effect.
\end{itemize}

The main problem with the differential technique is the contamination of isotropic strain channel to other anisotropic channels due to unequal strain transmission between two samples. This effect is inevitable as long as the samples are cut into bar-shape and glued along orthogonal directions \cite{shapiro2016measurement,kuo2016ubiquitous}. This effect becomes crucial when the elastoresistivity coefficient in the isotropic channel is much larger than the anisotropic channel. In this respect, both modified Montgomery and transverse techniques can perfectly eliminate this effect by measuring a single sample.

\textit{Elastocaloric effect measurement} - The elastocaloric effect was measured using an AC (dynamic) ECE technique with a homemade three-piezo-stack uniaxial strain cell, similar to the piezoelectric strain cell developed by Hicks et al. The sample was glued across the gap of the sample mounts of the strain cell. The temperature of the sample was measured at the center of the sample using a Chromel-AuFe0.07\% ($25~\mu m$ wire diameter) or type-E thermocouple for lower or higher temperature range, respectively. To apply a dynamical strain together with a static offset, AC and DC voltages were applied to the piezostacks using a TEGAM 2350 high-voltage amplifier sourced by a Stanford Research SR860 Lock-in amplifier and Keithley 2450 voltage supply, respectively. The signals from the sample and thermocouple were amplified by the Stanford Research SR554 preamplifier and measured by SR860. We present the thermal transfer function and phase as a function of frequency in Fig. S5 in the Supplementary Materials. Accordingly, we set the frequency 25.5 Hz to achieve the quasi-adiabatic condition in our measurements. The details of the AC ECE technique can be found in Refs. \cite{ikeda2019ac, ikeda2021elastocaloric, straquadine2020frequency}.

\section*{Data Availability}
Source data are available for this paper. All other data that support the plots within this paper and other findings of this study are available from the corresponding author upon reasonable request.

\section*{Acknowledgments}
We thank L. Balents, R. M. Fernandes, S. D. Wilson, Z. Wang, I. Zeljkovic and L. Zhao for helpful discussions. This work was mainly supported by NSF MRSEC at UW (DMR-2308979). Material synthesis at University of Washington was partially supported as part of Programmable Quantum Materials, an Energy Frontier Research Center funded by the U.S. Department of Energy (DOE), Office of Science, Basic Energy Sciences (BES), under award no. DE-SC0019443. J.-H.C. also acknowledges support from the David and Lucile Packard Foundation and the support from the State of Washington funded Clean Energy Institute. The work at Beijing Institute of Technology was supported by the National Science Foundation of China (NSFC) (grant No. 92065109), Beijing Natural Science Foundation (grant nos Z210006 and Z190006). Z.W. thanks the Analysis and Testing Center at BIT for assistance in facility support. A portion of this work was performed at the National High Magnetic Field Laboratory, which is supported by the National Science Foundation Cooperative Agreement No. DMR-1644779 and the State of Florida. \\

\section*{Author contributions}
J.-H.C. supervised the project. Z.L. performed the measurements with the help of Y.S., Q.J., E.R., J.D., C.H. and Y.Z. The single crystals were provided by J.L., Z.W. and Y.Y. High magnetic field measurements were supported by D.G. Z.L. and J.-H.C. analyzed the data and wrote the paper with comments from all coauthors on the paper.

\section*{Competing interests}
The authors declare no competing interests.

\section*{Additional Information}
Correspondence and requests for materials should be addressed to Z.L. and J.-H.C.

\bibliographystyle{naturemag}
\bibliography{kagome_CVS}

\begin{thebibliography}{10}
\expandafter\ifx\csname url\endcsname\relax
  \def\url#1{\texttt{#1}}\fi
\expandafter\ifx\csname urlprefix\endcsname\relax\def\urlprefix{URL }\fi
\providecommand{\bibinfo}[2]{#2}
\providecommand{\eprint}[2][]{\url{#2}}

\bibitem{neupert_charge_2022}
\bibinfo{author}{Neupert, T.}, \bibinfo{author}{Denner, M.~M.}, \bibinfo{author}{Yin, J.~X.}, \bibinfo{author}{Thomale, R.} \& \bibinfo{author}{Hasan, M.~Z.}
\newblock \bibinfo{title}{Charge order and superconductivity in kagome materials}.
\newblock \emph{\bibinfo{journal}{Nature Physics}} \textbf{\bibinfo{volume}{18}}, \bibinfo{pages}{137--143} (\bibinfo{year}{2022}).

\bibitem{guo2009topological}
\bibinfo{author}{Guo, H.-M.} \& \bibinfo{author}{Franz, M.}
\newblock \bibinfo{title}{Topological insulator on the kagome lattice}.
\newblock \emph{\bibinfo{journal}{Physical Review B}} \textbf{\bibinfo{volume}{80}}, \bibinfo{pages}{113102} (\bibinfo{year}{2009}).

\bibitem{liu_giant_2018}
\bibinfo{author}{Liu, E.} \emph{et~al.}
\newblock \bibinfo{title}{Giant anomalous {Hall} effect in a ferromagnetic kagome-lattice semimetal}.
\newblock \emph{\bibinfo{journal}{Nature Physics}} \textbf{\bibinfo{volume}{14}}, \bibinfo{pages}{1125--1131} (\bibinfo{year}{2018}).

\bibitem{ye2018massive}
\bibinfo{author}{Ye, L.} \emph{et~al.}
\newblock \bibinfo{title}{Massive {Dirac} fermions in a ferromagnetic kagome metal}.
\newblock \emph{\bibinfo{journal}{Nature}} \textbf{\bibinfo{volume}{555}}, \bibinfo{pages}{638--642} (\bibinfo{year}{2018}).

\bibitem{arachchige2022charge}
\bibinfo{author}{Arachchige, H. W.~S.} \emph{et~al.}
\newblock \bibinfo{title}{Charge density wave in kagome lattice intermetallic {ScV$_6$Sn$_6$}}.
\newblock \emph{\bibinfo{journal}{Physical Review Letters}} \textbf{\bibinfo{volume}{129}}, \bibinfo{pages}{216402} (\bibinfo{year}{2022}).

\bibitem{teng2022discovery}
\bibinfo{author}{Teng, X.} \emph{et~al.}
\newblock \bibinfo{title}{Discovery of charge density wave in a kagome lattice antiferromagnet}.
\newblock \emph{\bibinfo{journal}{Nature}} \textbf{\bibinfo{volume}{609}}, \bibinfo{pages}{490--495} (\bibinfo{year}{2022}).

\bibitem{ortiz_cs_2020}
\bibinfo{author}{Ortiz, B.~R.} \emph{et~al.}
\newblock \bibinfo{title}{{CsV$_3$Sb$_5$}: {A} {Z$_2$} {Topological} {Kagome} {Metal} with a {Superconducting} {Ground} {State}}.
\newblock \emph{\bibinfo{journal}{Physical Review Letters}} \textbf{\bibinfo{volume}{125}}, \bibinfo{pages}{247002} (\bibinfo{year}{2020}).

\bibitem{ortiz_new_2019}
\bibinfo{author}{Ortiz, B.~R.} \emph{et~al.}
\newblock \bibinfo{title}{New kagome prototype materials: {Discovery} of {KV$_3$Sb$_5$}, {RbV$_3$Sb$_5$}, and {CsV$_3$Sb$_5$}}.
\newblock \emph{\bibinfo{journal}{Physical Review Materials}} \textbf{\bibinfo{volume}{3}}, \bibinfo{pages}{94407} (\bibinfo{year}{2019}).

\bibitem{liang_three-dimensional_2021}
\bibinfo{author}{Liang, Z.} \emph{et~al.}
\newblock \bibinfo{title}{Three-{Dimensional} {Charge} {Density} {Wave} and {Surface}-{Dependent} {Vortex}-{Core} {States} in a {Kagome} {Superconductor} {CsV$_3$Sb$_5$}}.
\newblock \emph{\bibinfo{journal}{Physical Review X}} \textbf{\bibinfo{volume}{11}}, \bibinfo{pages}{031026} (\bibinfo{year}{2021}).

\bibitem{zhao2021cascade}
\bibinfo{author}{Zhao, H.} \emph{et~al.}
\newblock \bibinfo{title}{Cascade of correlated electron states in the kagome superconductor {CsV$_3$Sb$_5$}}.
\newblock \emph{\bibinfo{journal}{Nature}} \textbf{\bibinfo{volume}{599}}, \bibinfo{pages}{216--221} (\bibinfo{year}{2021}).

\bibitem{jiang2021unconventional}
\bibinfo{author}{Jiang, Y.-X.} \emph{et~al.}
\newblock \bibinfo{title}{Unconventional chiral charge order in kagome superconductor {KV$_3$Sb$_5$}}.
\newblock \emph{\bibinfo{journal}{Nature materials}} \textbf{\bibinfo{volume}{20}}, \bibinfo{pages}{1353--1357} (\bibinfo{year}{2021}).

\bibitem{li2022discovery}
\bibinfo{author}{Li, H.} \emph{et~al.}
\newblock \bibinfo{title}{Discovery of conjoined charge density waves in the kagome superconductor {CsV}$_{3}${Sb}$_{5}$}.
\newblock \emph{\bibinfo{journal}{Nature Communications}} \textbf{\bibinfo{volume}{13}}, \bibinfo{pages}{6348} (\bibinfo{year}{2022}).

\bibitem{xiao_coexistence_2023}
\bibinfo{author}{Xiao, Q.} \emph{et~al.}
\newblock \bibinfo{title}{Coexistence of multiple stacking charge density waves in kagome superconductor {CsV$_3$Sb$_5$}}.
\newblock \emph{\bibinfo{journal}{Physical Review Research}} \textbf{\bibinfo{volume}{5}}, \bibinfo{pages}{L012032} (\bibinfo{year}{2023}).

\bibitem{jiang_kagome_2023}
\bibinfo{author}{Jiang, K.} \emph{et~al.}
\newblock \bibinfo{title}{Kagome superconductors {AV}$_3${Sb}$_5$ ({A} = {K}, {Rb}, {Cs})}.
\newblock \emph{\bibinfo{journal}{National Science Review}} \textbf{\bibinfo{volume}{10}}, \bibinfo{pages}{nwac199} (\bibinfo{year}{2023}).

\bibitem{li_observation_2021}
\bibinfo{author}{Li, H.} \emph{et~al.}
\newblock \bibinfo{title}{Observation of unconventional charge density wave without acoustic phonon anomaly in kagome superconductors {$A$V$_3$Sb$_5$} ({A}={Rb}, {Cs})}.
\newblock \emph{\bibinfo{journal}{Physical Review X}} \textbf{\bibinfo{volume}{11}}, \bibinfo{pages}{1--9} (\bibinfo{year}{2021}).

\bibitem{xie_electron-phonon_2022}
\bibinfo{author}{Xie, Y.} \emph{et~al.}
\newblock \bibinfo{title}{Electron-phonon coupling in the charge density wave state of {CsV$_3$Sb$_5$}}.
\newblock \emph{\bibinfo{journal}{Physical Review B}} \textbf{\bibinfo{volume}{105}}, \bibinfo{pages}{L140501} (\bibinfo{year}{2022}).

\bibitem{christensen_theory_2021}
\bibinfo{author}{Christensen, M.~H.}, \bibinfo{author}{Birol, T.}, \bibinfo{author}{Andersen, B.~M.} \& \bibinfo{author}{Fernandes, R.~M.}
\newblock \bibinfo{title}{Theory of the charge density wave in {AV$_3$Sb$_5$} kagome metals}.
\newblock \emph{\bibinfo{journal}{Physical Review B}} \textbf{\bibinfo{volume}{104}}, \bibinfo{pages}{214513} (\bibinfo{year}{2021}).

\bibitem{wilson2024av}
\bibinfo{author}{Wilson, S.~D.} \& \bibinfo{author}{Ortiz, B.~R.}
\newblock \bibinfo{title}{{$A$V$_3$Sb$_5$} kagome superconductors}.
\newblock \emph{\bibinfo{journal}{Nature Reviews Materials}} \textbf{\bibinfo{volume}{9}}, \bibinfo{pages}{420–432} (\bibinfo{year}{2024}).

\bibitem{yang_giant_2020}
\bibinfo{author}{Yang, S.~Y.} \emph{et~al.}
\newblock \bibinfo{title}{Giant, unconventional anomalous {Hall} effect in the metallic frustrated magnet candidate, {KV$_3$Sb$_5$}}.
\newblock \emph{\bibinfo{journal}{Science Advances}} \textbf{\bibinfo{volume}{6}}, \bibinfo{pages}{1--8} (\bibinfo{year}{2020}).

\bibitem{yu_concurrence_2021}
\bibinfo{author}{Yu, F.~H.} \emph{et~al.}
\newblock \bibinfo{title}{Concurrence of anomalous {Hall} effect and charge density wave in a superconducting topological kagome metal}.
\newblock \emph{\bibinfo{journal}{Phys. Rev. B}} \textbf{\bibinfo{volume}{104}}, \bibinfo{pages}{L041103} (\bibinfo{year}{2021}).

\bibitem{yu_evidence_2021}
\bibinfo{author}{Yu, L.} \emph{et~al.}
\newblock \bibinfo{title}{Evidence of a hidden flux phase in the topological kagome metal {CsV$_3$Sb$_5$}}.
\newblock \emph{\bibinfo{journal}{arXiv preprint arXiv:2107.10714}}  (\bibinfo{year}{2021}).

\bibitem{xu_three-state_2022}
\bibinfo{author}{Xu, Y.} \emph{et~al.}
\newblock \bibinfo{title}{Three-state nematicity and magneto-optical {Kerr} effect in the charge density waves in kagome superconductors}.
\newblock \emph{\bibinfo{journal}{Nature physics}} \textbf{\bibinfo{volume}{18}}, \bibinfo{pages}{1470--1475} (\bibinfo{year}{2022}).

\bibitem{mielke_time-reversal_2022}
\bibinfo{author}{Mielke, C.} \emph{et~al.}
\newblock \bibinfo{title}{Time-reversal symmetry-breaking charge order in a kagome superconductor}.
\newblock \emph{\bibinfo{journal}{Nature}} \textbf{\bibinfo{volume}{602}}, \bibinfo{pages}{245--250} (\bibinfo{year}{2022}).

\bibitem{chen2022anomalous}
\bibinfo{author}{Chen, D.} \emph{et~al.}
\newblock \bibinfo{title}{Anomalous thermoelectric effects and quantum oscillations in the kagome metal {CsV$_3$Sb$_5$}}.
\newblock \emph{\bibinfo{journal}{Physical Review B}} \textbf{\bibinfo{volume}{105}}, \bibinfo{pages}{L201109} (\bibinfo{year}{2022}).

\bibitem{guo2022switchable}
\bibinfo{author}{Guo, C.} \emph{et~al.}
\newblock \bibinfo{title}{Switchable chiral transport in charge-ordered kagome metal {CsV$_3$Sb$_5$}}.
\newblock \emph{\bibinfo{journal}{Nature}} \textbf{\bibinfo{volume}{611}}, \bibinfo{pages}{461--466} (\bibinfo{year}{2022}).

\bibitem{xing2024optical}
\bibinfo{author}{Xing, Y.} \emph{et~al.}
\newblock \bibinfo{title}{Optical manipulation of the charge-density-wave state in {RbV$_3$Sb$_5$}}.
\newblock \emph{\bibinfo{journal}{Nature}}  (\bibinfo{year}{2024}).

\bibitem{feng_low-energy_2021}
\bibinfo{author}{Feng, X.}, \bibinfo{author}{Zhang, Y.}, \bibinfo{author}{Jiang, K.} \& \bibinfo{author}{Hu, J.}
\newblock \bibinfo{title}{Low-energy effective theory and symmetry classification of flux phases on the kagome lattice}.
\newblock \emph{\bibinfo{journal}{Phys. Rev. B}} \textbf{\bibinfo{volume}{104}}, \bibinfo{pages}{165136} (\bibinfo{year}{2021}).

\bibitem{christensen_loop_2022-1}
\bibinfo{author}{Christensen, M.~H.}, \bibinfo{author}{Birol, T.}, \bibinfo{author}{Andersen, B.~M.} \& \bibinfo{author}{Fernandes, R.~M.}
\newblock \bibinfo{title}{Loop currents in {AV$_3$Sb$_5$} kagome metals: {Multipolar} and toroidal magnetic orders}.
\newblock \emph{\bibinfo{journal}{Physical Review B}} \textbf{\bibinfo{volume}{106}}, \bibinfo{pages}{144504} (\bibinfo{year}{2022}).

\bibitem{saykin2022high}
\bibinfo{author}{Saykin, D.~R.} \emph{et~al.}
\newblock \bibinfo{title}{High resolution polar kerr effect studies of {CsV$_3$Sb$_5$}: Tests for time-reversal symmetry breaking below the charge-order transition}.
\newblock \emph{\bibinfo{journal}{Phys. Rev. Lett.}} \textbf{\bibinfo{volume}{131}}, \bibinfo{pages}{016901} (\bibinfo{year}{2023}).

\bibitem{wang2024resolving}
\bibinfo{author}{Wang, J.}, \bibinfo{author}{Farhang, C.}, \bibinfo{author}{Ortiz, B.~R.}, \bibinfo{author}{Wilson, S.~D.} \& \bibinfo{author}{Xia, J.}
\newblock \bibinfo{title}{Resolving the discrepancy between moke measurements at 1550-nm wavelength on kagome metal {CsV$_3$Sb$_5$}}.
\newblock \emph{\bibinfo{journal}{Physical Review Materials}} \textbf{\bibinfo{volume}{8}}, \bibinfo{pages}{014202} (\bibinfo{year}{2024}).

\bibitem{xiang_twofold_2021}
\bibinfo{author}{Xiang, Y.} \emph{et~al.}
\newblock \bibinfo{title}{Twofold symmetry of c-axis resistivity in topological kagome superconductor {CsV$_3$Sb$_5$} with in-plane rotating magnetic field}.
\newblock \emph{\bibinfo{journal}{Nature communications}} \textbf{\bibinfo{volume}{12}}, \bibinfo{pages}{6727} (\bibinfo{year}{2021}).

\bibitem{li_rotation_2022}
\bibinfo{author}{Li, H.} \emph{et~al.}
\newblock \bibinfo{title}{Rotation symmetry breaking in the normal state of a kagome superconductor {KV$_3$Sb$_5$}}.
\newblock \emph{\bibinfo{journal}{Nature Physics}} \textbf{\bibinfo{volume}{18}}, \bibinfo{pages}{265--270} (\bibinfo{year}{2022}).

\bibitem{li_unidirectional_2023}
\bibinfo{author}{Li, H.} \emph{et~al.}
\newblock \bibinfo{title}{Unidirectional coherent quasiparticles in the high-temperature rotational symmetry broken phase of {AV$_3$Sb$_5$} kagome superconductors}.
\newblock \emph{\bibinfo{journal}{Nature Physics}} \bibinfo{pages}{1--7} (\bibinfo{year}{2023}).

\bibitem{wulferding_emergent_2022}
\bibinfo{author}{Wulferding, D.} \emph{et~al.}
\newblock \bibinfo{title}{Emergent nematicity and intrinsic versus extrinsic electronic scattering processes in the kagome metal {CsV$_3$Sb$_5$}}.
\newblock \emph{\bibinfo{journal}{Physical Review Research}} \textbf{\bibinfo{volume}{4}}, \bibinfo{pages}{023215} (\bibinfo{year}{2022}).

\bibitem{wu_charge_2022}
\bibinfo{author}{Wu, S.} \emph{et~al.}
\newblock \bibinfo{title}{Charge density wave order in the kagome metal {AV$_3$Sb$_5$} ( {A} = {Cs} , {Rb} , {K} )}.
\newblock \emph{\bibinfo{journal}{Physical Review B}} \textbf{\bibinfo{volume}{105}}, \bibinfo{pages}{155106} (\bibinfo{year}{2022}).

\bibitem{wu_simultaneous_2022}
\bibinfo{author}{Wu, Q.} \emph{et~al.}
\newblock \bibinfo{title}{Simultaneous formation of two-fold rotation symmetry with charge order in the kagome superconductor {CsV$_3$Sb$_5$} by optical polarization rotation measurement}.
\newblock \emph{\bibinfo{journal}{Physical Review B}} \textbf{\bibinfo{volume}{106}}, \bibinfo{pages}{205109} (\bibinfo{year}{2022}).

\bibitem{jiang_observation_2023}
\bibinfo{author}{Jiang, Z.} \emph{et~al.}
\newblock \bibinfo{title}{Observation of electronic nematicity driven by the three-dimensional charge density wave in kagome lattice {KV$_3$Sb$_5$}}.
\newblock \emph{\bibinfo{journal}{Nano Letters}} \textbf{\bibinfo{volume}{23}}, \bibinfo{pages}{5625--5633} (\bibinfo{year}{2023}).

\bibitem{luo2022electronic}
\bibinfo{author}{Luo, H.} \emph{et~al.}
\newblock \bibinfo{title}{Electronic nature of charge density wave and electron-phonon coupling in kagome superconductor {KV}$_{3}${Sb}$_{5}$}.
\newblock \emph{\bibinfo{journal}{Nature communications}} \textbf{\bibinfo{volume}{13}}, \bibinfo{pages}{273} (\bibinfo{year}{2022}).

\bibitem{kang2023charge}
\bibinfo{author}{Kang, M.} \emph{et~al.}
\newblock \bibinfo{title}{Charge order landscape and competition with superconductivity in kagome metals}.
\newblock \emph{\bibinfo{journal}{Nature Materials}} \textbf{\bibinfo{volume}{22}}, \bibinfo{pages}{186--193} (\bibinfo{year}{2023}).

\bibitem{nie_charge-density-wave-driven_2022}
\bibinfo{author}{Nie, L.} \emph{et~al.}
\newblock \bibinfo{title}{Charge-density-wave-driven electronic nematicity in a kagome superconductor}.
\newblock \emph{\bibinfo{journal}{Nature}} \textbf{\bibinfo{volume}{604}}, \bibinfo{pages}{59--64} (\bibinfo{year}{2022}).

\bibitem{sur2023optimized}
\bibinfo{author}{Sur, Y.}, \bibinfo{author}{Kim, K.-T.}, \bibinfo{author}{Kim, S.} \& \bibinfo{author}{Kim, K.~H.}
\newblock \bibinfo{title}{Optimized superconductivity in the vicinity of a nematic quantum critical point in the kagome superconductor {Cs(V$_{1-x}$Ti$_x$)$_3$Sb$_5$}}.
\newblock \emph{\bibinfo{journal}{Nature Communications}} \textbf{\bibinfo{volume}{14}}, \bibinfo{pages}{3899} (\bibinfo{year}{2023}).

\bibitem{fradkin2010nematic}
\bibinfo{author}{Fradkin, E.}, \bibinfo{author}{Kivelson, S.~A.}, \bibinfo{author}{Lawler, M.~J.}, \bibinfo{author}{Eisenstein, J.~P.} \& \bibinfo{author}{Mackenzie, A.~P.}
\newblock \bibinfo{title}{Nematic fermi fluids in condensed matter physics}.
\newblock \emph{\bibinfo{journal}{Annu. Rev. Condens. Matter Phys.}} \textbf{\bibinfo{volume}{1}}, \bibinfo{pages}{153--178} (\bibinfo{year}{2010}).

\bibitem{fernandes2014drives}
\bibinfo{author}{Fernandes, R.}, \bibinfo{author}{Chubukov, A.} \& \bibinfo{author}{Schmalian, J.}
\newblock \bibinfo{title}{What drives nematic order in iron-based superconductors?}
\newblock \emph{\bibinfo{journal}{Nature physics}} \textbf{\bibinfo{volume}{10}}, \bibinfo{pages}{97--104} (\bibinfo{year}{2014}).

\bibitem{chu2012divergent}
\bibinfo{author}{Chu, J.-H.}, \bibinfo{author}{Kuo, H.-H.}, \bibinfo{author}{Analytis, J.~G.} \& \bibinfo{author}{Fisher, I.~R.}
\newblock \bibinfo{title}{Divergent nematic susceptibility in an iron arsenide superconductor}.
\newblock \emph{\bibinfo{journal}{Science}} \textbf{\bibinfo{volume}{337}}, \bibinfo{pages}{710--712} (\bibinfo{year}{2012}).

\bibitem{kuo2016ubiquitous}
\bibinfo{author}{Kuo, H.-H.}, \bibinfo{author}{Chu, J.-H.}, \bibinfo{author}{Palmstrom, J.~C.}, \bibinfo{author}{Kivelson, S.~A.} \& \bibinfo{author}{Fisher, I.~R.}
\newblock \bibinfo{title}{Ubiquitous signatures of nematic quantum criticality in optimally doped {F}e-based superconductors}.
\newblock \emph{\bibinfo{journal}{Science}} \textbf{\bibinfo{volume}{352}}, \bibinfo{pages}{958--962} (\bibinfo{year}{2016}).

\bibitem{liu2016nematic}
\bibinfo{author}{Liu, Z.} \emph{et~al.}
\newblock \bibinfo{title}{Nematic quantum critical fluctuations in {BaFe}$_{2-x}${Ni}$_x${As}$_2$}.
\newblock \emph{\bibinfo{journal}{Physical Review Letters}} \textbf{\bibinfo{volume}{117}}, \bibinfo{pages}{157002} (\bibinfo{year}{2016}).

\bibitem{gu2017unified}
\bibinfo{author}{Gu, Y.} \emph{et~al.}
\newblock \bibinfo{title}{Unified phase diagram for iron-based superconductors}.
\newblock \emph{\bibinfo{journal}{Physical Review Letters}} \textbf{\bibinfo{volume}{119}}, \bibinfo{pages}{157001} (\bibinfo{year}{2017}).

\bibitem{nie2014quenched}
\bibinfo{author}{Nie, L.}, \bibinfo{author}{Tarjus, G.} \& \bibinfo{author}{Kivelson, S.~A.}
\newblock \bibinfo{title}{Quenched disorder and vestigial nematicity in the pseudogap regime of the cuprates}.
\newblock \emph{\bibinfo{journal}{Proceedings of the National Academy of Sciences}} \textbf{\bibinfo{volume}{111}}, \bibinfo{pages}{7980--7985} (\bibinfo{year}{2014}).

\bibitem{fernandes2019intertwined}
\bibinfo{author}{Fernandes, R.~M.}, \bibinfo{author}{Orth, P.~P.} \& \bibinfo{author}{Schmalian, J.}
\newblock \bibinfo{title}{Intertwined vestigial order in quantum materials: Nematicity and beyond}.
\newblock \emph{\bibinfo{journal}{Annual Review of Condensed Matter Physics}} \textbf{\bibinfo{volume}{10}}, \bibinfo{pages}{133--154} (\bibinfo{year}{2019}).

\bibitem{bohmer2022nematicity}
\bibinfo{author}{B{\"o}hmer, A.~E.}, \bibinfo{author}{Chu, J.-H.}, \bibinfo{author}{Lederer, S.} \& \bibinfo{author}{Yi, M.}
\newblock \bibinfo{title}{Nematicity and nematic fluctuations in iron-based superconductors}.
\newblock \emph{\bibinfo{journal}{Nature Physics}} \textbf{\bibinfo{volume}{18}}, \bibinfo{pages}{1412--1419} (\bibinfo{year}{2022}).

\bibitem{park2021electronic}
\bibinfo{author}{Park, T.}, \bibinfo{author}{Ye, M.} \& \bibinfo{author}{Balents, L.}
\newblock \bibinfo{title}{Electronic instabilities of kagome metals: saddle points and landau theory}.
\newblock \emph{\bibinfo{journal}{Physical Review B}} \textbf{\bibinfo{volume}{104}}, \bibinfo{pages}{035142} (\bibinfo{year}{2021}).

\bibitem{grandi2023theory}
\bibinfo{author}{Grandi, F.}, \bibinfo{author}{Consiglio, A.}, \bibinfo{author}{Sentef, M.~A.}, \bibinfo{author}{Thomale, R.} \& \bibinfo{author}{Kennes, D.~M.}
\newblock \bibinfo{title}{Theory of nematic charge orders in kagome metals}.
\newblock \emph{\bibinfo{journal}{Physical Review B}} \textbf{\bibinfo{volume}{107}}, \bibinfo{pages}{155131} (\bibinfo{year}{2023}).

\bibitem{denner2021analysis}
\bibinfo{author}{Denner, M.~M.}, \bibinfo{author}{Thomale, R.} \& \bibinfo{author}{Neupert, T.}
\newblock \bibinfo{title}{Analysis of charge order in the kagome metal {AV$_3$Sb$_5$} ({A= K, Rb, Cs})}.
\newblock \emph{\bibinfo{journal}{Physical Review Letters}} \textbf{\bibinfo{volume}{127}}, \bibinfo{pages}{217601} (\bibinfo{year}{2021}).

\bibitem{tazai2022mechanism}
\bibinfo{author}{Tazai, R.}, \bibinfo{author}{Yamakawa, Y.}, \bibinfo{author}{Onari, S.} \& \bibinfo{author}{Kontani, H.}
\newblock \bibinfo{title}{Mechanism of exotic density-wave and beyond-migdal unconventional superconductivity in kagome metal av3sb5 (a= k, rb, cs)}.
\newblock \emph{\bibinfo{journal}{Science Advances}} \textbf{\bibinfo{volume}{8}}, \bibinfo{pages}{eabl4108} (\bibinfo{year}{2022}).

\bibitem{zhou2022chern}
\bibinfo{author}{Zhou, S.} \& \bibinfo{author}{Wang, Z.}
\newblock \bibinfo{title}{Chern fermi pocket, topological pair density wave, and charge-4e and charge-6e superconductivity in kagom{\'e} superconductors}.
\newblock \emph{\bibinfo{journal}{Nature Communications}} \textbf{\bibinfo{volume}{13}}, \bibinfo{pages}{7288} (\bibinfo{year}{2022}).

\bibitem{wiecki2020dominant}
\bibinfo{author}{Wiecki, P.} \emph{et~al.}
\newblock \bibinfo{title}{Dominant in-plane symmetric elastoresistance in {CsFe}$_2${As}$_2$}.
\newblock \emph{\bibinfo{journal}{Physical Review Letters}} \textbf{\bibinfo{volume}{125}}, \bibinfo{pages}{187001} (\bibinfo{year}{2020}).

\bibitem{wiecki2021emerging}
\bibinfo{author}{Wiecki, P.} \emph{et~al.}
\newblock \bibinfo{title}{Emerging symmetric strain response and weakening nematic fluctuations in strongly hole-doped iron-based superconductors}.
\newblock \emph{\bibinfo{journal}{Nature Communications}} \textbf{\bibinfo{volume}{12}}, \bibinfo{pages}{4824} (\bibinfo{year}{2021}).

\bibitem{rosenberg2019divergence}
\bibinfo{author}{Rosenberg, E.~W.}, \bibinfo{author}{Chu, J.-H.}, \bibinfo{author}{Ruff, J.~P.}, \bibinfo{author}{Hristov, A.~T.} \& \bibinfo{author}{Fisher, I.~R.}
\newblock \bibinfo{title}{Divergence of the quadrupole-strain susceptibility of the electronic nematic system {YbRu}$_2${Ge}$_2$}.
\newblock \emph{\bibinfo{journal}{Proceedings of the National Academy of Sciences}} \textbf{\bibinfo{volume}{116}}, \bibinfo{pages}{7232--7237} (\bibinfo{year}{2019}).

\bibitem{ye2023elastocaloric}
\bibinfo{author}{Ye, L.} \emph{et~al.}
\newblock \bibinfo{title}{Elastocaloric signatures of symmetric and antisymmetric strain-tuning of quadrupolar and magnetic phases in {DyB}$_2${C}$_2$}.
\newblock \emph{\bibinfo{journal}{Proceedings of the National Academy of Sciences}} \textbf{\bibinfo{volume}{120}}, \bibinfo{pages}{e2302800120} (\bibinfo{year}{2023}).

\bibitem{ortiz_fermi_2021}
\bibinfo{author}{Ortiz, B.~R.} \emph{et~al.}
\newblock \bibinfo{title}{Fermi surface mapping and the nature of charge density wave order in the kagome superconductor {CsV$_3$Sb$_5$}}.
\newblock \emph{\bibinfo{journal}{Physical Review X}} \textbf{\bibinfo{volume}{11}}, \bibinfo{pages}{41030} (\bibinfo{year}{2021}).

\bibitem{fu2021quantum}
\bibinfo{author}{Fu, Y.} \emph{et~al.}
\newblock \bibinfo{title}{Quantum transport evidence of topological band structures of kagome superconductor {CsV}$_{3}${Sb}$_{5}$}.
\newblock \emph{\bibinfo{journal}{Physical Review Letters}} \textbf{\bibinfo{volume}{127}}, \bibinfo{pages}{207002} (\bibinfo{year}{2021}).

\bibitem{fernandes2020nematicity}
\bibinfo{author}{Fernandes, R.~M.} \& \bibinfo{author}{Venderbos, J.~W.}
\newblock \bibinfo{title}{Nematicity with a twist: Rotational symmetry breaking in a moir{\'e} superlattice}.
\newblock \emph{\bibinfo{journal}{Science Advances}} \textbf{\bibinfo{volume}{6}}, \bibinfo{pages}{eaba8834} (\bibinfo{year}{2020}).

\bibitem{shapiro2016measurement}
\bibinfo{author}{Shapiro, M.}, \bibinfo{author}{Hristov, A.}, \bibinfo{author}{Palmstrom, J.}, \bibinfo{author}{Chu, J.-H.} \& \bibinfo{author}{Fisher, I.}
\newblock \bibinfo{title}{Measurement of the {B$_{1g}$} and {B$_{2g}$} components of the elastoresistivity tensor for tetragonal materials via transverse resistivity configurations}.
\newblock \emph{\bibinfo{journal}{Review of Scientific Instruments}} \textbf{\bibinfo{volume}{87}}, \bibinfo{pages}{063902} (\bibinfo{year}{2016}).

\bibitem{frachet2024colossal}
\bibinfo{author}{Frachet, M.} \emph{et~al.}
\newblock \bibinfo{title}{Colossal c-axis response and lack of rotational symmetry breaking within the kagome planes of the {CsV$_3$Sb$_5$} superconductor}.
\newblock \emph{\bibinfo{journal}{Physical Review Letters}} \textbf{\bibinfo{volume}{132}}, \bibinfo{pages}{186001} (\bibinfo{year}{2024}).

\bibitem{asaba2024evidence}
\bibinfo{author}{Asaba, T.} \emph{et~al.}
\newblock \bibinfo{title}{Evidence for an odd-parity nematic phase above the charge-density-wave transition in a kagome metal}.
\newblock \emph{\bibinfo{journal}{Nature Physics}} \textbf{\bibinfo{volume}{20}}, \bibinfo{pages}{40} (\bibinfo{year}{2024}).

\bibitem{inyushkin1998magnetic}
\bibinfo{author}{Inyushkin, A.}, \bibinfo{author}{Leicht, K.} \& \bibinfo{author}{Esquinazi, P.}
\newblock \bibinfo{title}{Magnetic field dependence of the sensitivity of a type {E} (chromel--constantan) thermocouple}.
\newblock \emph{\bibinfo{journal}{Cryogenics}} \textbf{\bibinfo{volume}{38}}, \bibinfo{pages}{299--304} (\bibinfo{year}{1998}).

\bibitem{ikeda2019ac}
\bibinfo{author}{Ikeda, M.~S.} \emph{et~al.}
\newblock \bibinfo{title}{{AC} elastocaloric effect as a probe for thermodynamic signatures of continuous phase transitions}.
\newblock \emph{\bibinfo{journal}{Review of Scientific Instruments}} \textbf{\bibinfo{volume}{90}}, \bibinfo{pages}{083902} (\bibinfo{year}{2019}).

\bibitem{ikeda2021elastocaloric}
\bibinfo{author}{Ikeda, M.~S.} \emph{et~al.}
\newblock \bibinfo{title}{Elastocaloric signature of nematic fluctuations}.
\newblock \emph{\bibinfo{journal}{Proceedings of the National Academy of Sciences}} \textbf{\bibinfo{volume}{118}}, \bibinfo{pages}{e2105911118} (\bibinfo{year}{2021}).

\bibitem{hristov2019elastoresistive}
\bibinfo{author}{Hristov, A.~T.}, \bibinfo{author}{Ikeda, M.~S.}, \bibinfo{author}{Palmstrom, J.~C.}, \bibinfo{author}{Walmsley, P.} \& \bibinfo{author}{Fisher, I.~R.}
\newblock \bibinfo{title}{Elastoresistive and elastocaloric anomalies at magnetic and electronic-nematic critical points}.
\newblock \emph{\bibinfo{journal}{Phys. Rev. B}} \textbf{\bibinfo{volume}{99}}, \bibinfo{pages}{100101} (\bibinfo{year}{2019}).

\bibitem{straquadine2020frequency}
\bibinfo{author}{Straquadine, J.}, \bibinfo{author}{Ikeda, M.} \& \bibinfo{author}{Fisher, I.}
\newblock \bibinfo{title}{Frequency-dependent sensitivity of {AC} elastocaloric effect measurements explored through analytical and numerical models}.
\newblock \emph{\bibinfo{journal}{Review of Scientific Instruments}} \textbf{\bibinfo{volume}{91}}, \bibinfo{pages}{083905} (\bibinfo{year}{2020}).

\bibitem{qian2021revealing}
\bibinfo{author}{Qian, T.} \emph{et~al.}
\newblock \bibinfo{title}{Revealing the competition between charge density wave and superconductivity in {CsV$_3$Sb$_5$} through uniaxial strain}.
\newblock \emph{\bibinfo{journal}{Physical Review B}} \textbf{\bibinfo{volume}{104}}, \bibinfo{pages}{144506} (\bibinfo{year}{2021}).

\bibitem{wu2022simultaneous}
\bibinfo{author}{Wu, Q.} \emph{et~al.}
\newblock \bibinfo{title}{Simultaneous formation of two-fold rotation symmetry with charge order in the kagome superconductor {CsV$_3$Sb$_5$} by optical polarization rotation measurement}.
\newblock \emph{\bibinfo{journal}{Physical Review B}} \textbf{\bibinfo{volume}{106}}, \bibinfo{pages}{205109} (\bibinfo{year}{2022}).

\bibitem{guo2024correlated}
\bibinfo{author}{Guo, C.} \emph{et~al.}
\newblock \bibinfo{title}{Correlated order at the tipping point in the kagome metal {CsV$_3$Sb$_5$}}.
\newblock \emph{\bibinfo{journal}{Nature Physics}} \textbf{\bibinfo{volume}{20}}, \bibinfo{pages}{579–584} (\bibinfo{year}{2024}).

\bibitem{wang_electronic_2021}
\bibinfo{author}{Wang, Z.} \emph{et~al.}
\newblock \bibinfo{title}{Electronic nature of chiral charge order in the kagome superconductor {CsV}$_{3}${Sb}$_{5}$}.
\newblock \emph{\bibinfo{journal}{Physical Review B}} \textbf{\bibinfo{volume}{104}}, \bibinfo{pages}{075148} (\bibinfo{year}{2021}).

\end{thebibliography}

\clearpage

\renewcommand{\thefigure}{S\arabic{figure}}
\renewcommand{\thesection}{S\arabic{section}}
\renewcommand{\thesubsection}{S\arabic{subsection}}
\renewcommand{\theequation}{S\arabic{equation}}
\renewcommand{\thetable}{S\arabic{table}}
\setcounter{figure}{0} 
\setcounter{equation}{0}
\appendix

\onecolumngrid

\end{document}